\documentstyle[12pt]{article}
\newcommand{\PSbox}[3]{\mbox{\rule{0in}{#3}\includegraphics{#1}\hspace{#2}}}
\newcommand{\mysection}[1]{\setcounter{equation}{0}\section{#1}}

\newcommand{\nc}{\newcommand}
\nc{\beq}{\begin{equation}} \nc{\eeq}{\end{equation}}
\nc{\beqa}{\begin{eqnarray}} \nc{\eeqa}{\end{eqnarray}}
\nc{\lsim}{\begin{array}{c}\,\sim\vspace{-21pt}\\< \end{array}}
\nc{\gsim}{\begin{array}{c}\sim\vspace{-21pt}\\> \end{array}}
\newfont\figfont{cmr7 scaled 1200}
\begin{document}

\begin{titlepage}
\begin{center}
{\hbox to\hsize{hep-ph/9512278 \hfill  MIT-CTP-2478}}

\bigskip

\bigskip

{\Large \bf   Phenomenological Constraints on the Higgs
as Pseudo-Goldstone Boson Mechanism in Supersymmetric GUT Theories
\footnotemark[1]      } \\

\bigskip

\bigskip

{\bf Csaba Cs\'aki and  Lisa Randall}\footnotemark[2]\\

\smallskip

{ \small \it Center for Theoretical Physics

Laboratory for Nuclear Science and Department of Physics

Massachusetts Institute of Technology

Cambridge, MA 02139, USA }

 \bigskip

\vspace{2cm}

{\bf Abstract}\\[-0.05in]
\end{center}

There are few robust solutions to the doublet-triplet splitting problem
in supersymmetric GUT theories.  One of the more promising solutions
is the Higgs as pseudo-Goldstone boson mechanism.  In its
minimal implementation, such a solution places an additional
restriction on the parameter space of the minimal supersymmetric
standard model.  A testable consequence of this constraint is
an equation for $\tan \beta$. We present this restriction
and study its solutions in order to constrain the allowed
parameter space. Thus the assumptions on the
GUT scale Higgs sector should
yield testable predictions for weak scale
physics. If the SUSY parameters are measured
then it should be possible to check the predictions, yielding
insight into GUT scale physics.

\bigskip

\footnotetext[1]{Supported in part
by DOE under cooperative agreement \#DE-FC02-94ER40818.}
\footnotetext[2]{NSF Young Investigator Award, Alfred P.~Sloan
Foundation Fellowship, DOE Outstanding Junior
Investigator Award. }

\end{titlepage}

\mysection{Introduction}

The unification of couplings in the MSSM is strong evidence for
supersymmetric grand unified theories (SUSY GUT's).
However a credible SUSY GUT  should incorporate
a solution to  the
doublet-triplet splitting problem and the
associated $\mu$-problem.

The Higgs as pseudo-Goldstone boson (PGB) scheme elegantly solves both
 problems \cite{Jap}.
In this solution the Higgs fields are light
(of the order of the weak scale) because an additional global symmetry
is spontaneously broken implying the presence of some massless Goldstone
bosons.  By suitable choice of the gauge and global symmetry
groups one can achieve  a situation in which the pseudo-Goldstone
bosons are exactly
one pair of SU(2) doublets to be identified with the Higgs fields of the
MSSM [1-9].
At the same time this mechanism naturally generates a $\mu$-term.

In this paper we investigate the restrictions that
the assumption that the
$\mu$-term is generated purely by the Higgs as PGB mechanism
places on the low energy physics.
Because of an additional constraint on the
Higgs sector $\tan \beta$  is not an independent parameter.  We
impose this constraint and check whether a suitable standard model
minimum exists. This will result in an equation for $\tan \beta$
which is given in eq. \ref{theeq2}.

We solve this equation numerically and give the resulting
range of $\tan \beta$. The parameter range where this
equation can be satisfied is also displayed.

Thus we will show how our assumptions on the GUT physics
together with our
present knowledge of weak scale phenomenology
constrain the parameters of weak scale supersymmetry.
This way we could get information on the GUT scale
physics if the MSSM parameters are ultimately
measured in future colliders.

The paper is organized as follows: in Section 2 we review
how a $\mu$-term is generated in the Higgs
as PGB scheme. Section 3 contains the analysis of the extra constraint -
which reduces the number of independent MSSM parameters by one. In Section 4
we
discuss our numerical results and the resulting allowed parameter space of
the MSSM.
The consequences of these results for realistic SUSY GUT models
is discussed in Section 5.
We conclude in Section 6.

\mysection{The $\mu$-term from the Higgs as PGB scheme}

In the Higgs as PGB solution to the doublet-triplet splitting problem
one assumes that there is an additional global symmetry which, after
spontaneous breaking, ensures the lightness of the Higgs particles \cite{Jap}.
For example, in the originally proposed model the SU(5) gauge group is
enlarged to an SU(6) global symmetry containing the gauged SU(5). While
the SU(6) breaks to SU(4)$\otimes $SU(2)$\otimes$U(1), the gauged
SU(5) breaks to the SM group SU(3)$\otimes$SU(2)$\otimes$U(1), leaving one
pair of SU(2) doublets as uneaten Goldstone bosons
 \cite{Jap,Ans}.

However only one of the scalars in the chiral superfields is a genuine
Goldstone boson;
the other scalar is massless only on account of supersymmetry. Thus it is
not surprising that   the soft breaking terms (which do
preserve the extra global symmetry) generate  mass terms for these
non-Goldstone boson
light fields and a supersymmetric $\mu$-term for the Higgs fields.

In this section we will explicitly demonstrate how this mechanism works in
the above mentioned SU(5) model with SU(6) global symmetry.
Although this model is not the most aesthetic in
that it requires tuning the SU(5) couplings, the simplicity
of this model makes it a good  example to illustrate the generic features of
the Higgs as PGB mechanism. We will also see in section 5 that this model
can be an effective theory for some energy range for
the more realistic model of refs. \cite{Zur,Bar2,Bar3,Zur2,BCR}.
All the results in this section apply more generally
as has been shown  in  ref. \cite{Jap}.

The Higgs sector of  this model consists of  one adjoint of SU(6), denoted by
$\Phi$,
\begin{equation}
\Phi=24+5+\bar{5}+1=\Sigma +H+\bar{H}+S
\end{equation}
under SU(5). The explicit realization of this decomposition is
given in the following way:
\begin{equation}
\Phi=\left( \begin{array}{cc} -\frac{5}{\sqrt{30}}S & H \\
\bar{H} & \Sigma + \frac{1}{\sqrt{30}}S \end{array} \right).
\end{equation}
The superpotential is given by
\begin{equation}
W(\Phi)=\frac{1}{2} M {\rm Tr} \Phi^2 +\frac{1}{3} \lambda {\rm Tr} \Phi^3.
\end{equation}
The VEV is given by
\begin{equation}
\langle \Phi \rangle = \frac{M}{\lambda} \left( \begin{array}{cccccc}
1 & & & & & \\ & 1 & & & & \\ & & 1 & & & \\ & & & 1 & & \\ & & & & -2 & \\
& & & & & -2 \end{array} \right),
\end{equation}
which breaks the gauged SU(5) to SU(3)$\otimes $SU(2)$\otimes$U(1), and leaves
one pair of Goldstone bosons uneaten, which can be identified as the SU(2)
doublets
in $H$ and $\bar{H}$. The scalar potential after the inclusion of the soft
breaking terms is given by
\begin{equation}
\label{softpot}
V(\Phi )={\rm Tr}\left| M\Phi +\lambda (\Phi^2-\frac{1}{6}{\rm Tr} \Phi^2)
\right|^2+
\left( A_{\Phi}
\lambda \frac{1}{3} {\rm Tr} \Phi^3+B_{\Phi}M \frac{1}{2}
{\rm Tr}\Phi^2 +h.c.\right) +
m_0^2 {\rm Tr} |\Phi|^2.
\end{equation}
Now the $\Phi$ VEV is shifted to
\begin{equation}
\langle \Phi \rangle = \frac{1}{\lambda} \left[ M+(A_{\Phi}-
B_{\Phi})+\frac{1}{M} (3A_{\Phi}B_{\Phi}-A_{\Phi}^2-
2B_{\Phi}^2-m_0^2)+{\cal O}\left( \frac{1}{M^2}\right) \right].
\end{equation}
After substituting $\langle \Phi \rangle$ into the potential
one finds a mass term for the doublets in
$H,\bar{H}$ (denoted by $h_u,h_d$)
\begin{equation}
\left| \frac{h_u-h_d^{\dagger}}{\sqrt{2}} \right|^2 \left[ 2m_0^2
+2(A_{\Phi}-B_{\Phi})^2\right],
\end{equation}
while at the same time the shift in the $\Phi$ VEV generates a mass term
for the higgsinos $\tilde{h_u}$ and $\tilde{h_d}$ of the form
\begin{equation}
(A_{\Phi}-B_{\Phi}) \tilde{h_u}^c\tilde{h_d}+h.c..
\end{equation}
Thus, as it was shown in general in ref. \cite{Jap}, there is a supersymmetric
$\mu$-term generated by this mechanism, which is usually a model dependent
function of the soft breaking parameters (in the above described
model $\mu = A_{\Phi}-B_{\Phi}$).
The other important lesson from this example is that the
combination $1/\sqrt{2}(h_u+h_d^{\dagger})$ is the genuine Goldstone
boson with no mass
term at the GUT scale even after inclusion of the soft breaking terms
\cite{Jap}. Such a
mass term is generated only by the explicit symmetry breaking loop
corrections.

Thus, the general conclusion is that the Higgs as PGB mechanism generates
the following mass term for the MSSM Higgs fields at the GUT scale:
\begin{equation}
\label{pgbpot}
V(h_u,h_d)|_{M_{GUT}}=(m_0^2+\mu^2) |h_u-h_d^{\dagger}|^2+\;
\mbox{D-terms},
\end{equation}
where $m_0$ is the soft breaking mass parameter introduced in
eq. \ref{softpot}, while $\mu$ is a function of
 all soft breaking parameters.

The general Higgs potential in the MSSM is given by
\begin{eqnarray}
V(h_u,h_d)|_{\Lambda}= m_1^2(\Lambda ) h_d^{\dagger}h_d+m_2^2(\Lambda )
h_u^{\dagger} h_u+
m_3^2(\Lambda )(h_uh_d+ h.c.)+\; \mbox{D-terms}
\end{eqnarray}
where we have explicitly displayed the scale ($\Lambda$)
dependence of the parameters and where
\begin{eqnarray}
& & m_1^2 (\Lambda )=m_d^2(\Lambda )+\mu^2(\Lambda ), \nonumber \\
& & m_2^2 (\Lambda )=m_u^2(\Lambda )+\mu^2(\Lambda ), \nonumber \\
& & m_3^2 (\Lambda )=B(\Lambda )\mu (\Lambda ),
\end{eqnarray}
$m_u^2,m_d^2$ are the running soft breaking mass terms for the up- and
down-type Higgses, $\mu (\Lambda)$ is the running $\mu$-parameter, while
$B(\Lambda )$ is the running soft breaking parameter corresponding to the
$\mu$-term. From eq. \ref{pgbpot} one can see that the specific $\mu$-term
generated by the Higgs as PGB mechanism requires that the boundary condition
\begin{equation}
\label{constr}
m_d^2(M_G)+\mu^2(M_G)=m_u^2(M_G)+\mu^2(M_G)=-B(M_G)\mu (M_G)
\end{equation}
is satisfied.
If one assumes universal soft breaking terms at the GUT-scale,
then the first of the equations is automatically satisfied, and
we have
one additional constraint equation.

As noted already in refs. \cite{Jap,Ans,Jap2,Bar2},
eq. \ref{constr} means that the number of free
parameters in the MSSM is reduced by one. However we have not seen in any of
 the previous analyses an explicit  determination
of the results of this additional restriction.

Already the authors of the first papers on
this subject noted the importance of eq.
\ref{constr}, and analyzed its consequences. However in these
papers \cite{Ans,Jap2}
a very specific form of the soft breaking terms  was assumed and therefore
the consequences were not sufficiently general. The authors of
ref. \cite{Bar2} also provide an analysis of the
constraint of eq. \ref{constr}.
However their method is not described in sufficient
detail for us to compare the results.

\mysection{Analysis of the constraint arising from the Higgs as PGB
mechanism}
As we saw in the previous section, the specific form of the $\mu$-term
generated by the Higgs as PGB mechanism implies the following constraint
on the running mass parameters:
\begin{equation}
\label{constr2}
m_d^2(M)+\mu^2(M)=m_u^2(M)+\mu^2(M)=-B(M)\mu (M)
\end{equation}
where $M$ is the scale  where the additional global symmetry
which gives rise to the light Higgses is broken.
We make the following assumptions:

A. The $\mu$-term is generated purely by the Higgs as PGB
mechanism implying the constraint \ref{constr2}.

B. The form of the constraint \ref{constr2} remains
valid at the GUT-scale.\footnote{Our results do not
change significantly when allowing running between
$M_P$ and $M_G$. See the end of Section 5.}

\noindent Usually the number of free parameters in the
MSSM (assuming universal
soft breaking terms at the GUT scale and gaugino unification) is 5+1, where
the 5+1 are:

1. $m_0$ - the universal soft breaking mass term,

2. $A_0$ -  the trilinear soft breaking term,

3. $M_{1/2}$ - the gaugino mass,

4. $\lambda_t$ - the top Yukawa coupling,

5. $\tan \beta$ - the ratio of Higgs VEV's,

\noindent and the extra parameter is the sign of the $\mu$ parameter.

\noindent We can see that this set does not contain either $\mu$ or $B$, since
they are determined (at the weak scale) from the requirement of electroweak
symmetry breaking (see e.g. \cite{Diego}):

\begin{eqnarray}
\label{ewbr}
&& \mu^2(M_Z)=\frac{\bar{m}_d^2-\bar{m}_u^2\tan^2 \beta}{\tan^2 \beta -1}-
\frac{1}{2} M_Z^2, \nonumber \\
&& B(M_Z)\mu (M_Z)=\frac{1}{2} \sin 2\beta (\bar{m}_1^2+\bar{m}_2^2),
\end{eqnarray}
where $\bar{m}_d^2,\bar{m}_u^2,\bar{m}_1^2,\bar{m}_2^2$ are the 1-loop
corrected values of the above defined soft breaking mass parameters
evaluated at the weak scale.

This means that the constraint eq. \ref{constr2} will determine one additional
parameter of the 5, but in a   nontrivial way. In our analysis
we choose $m_0,A_0,M_{1/2},\lambda_t$ and the sign of    $\mu$ to be the
independent parameters and $\tan \beta$ will be evaluated in the following
way: given the 4 input parameters, one can calculate the soft breaking
mass terms at the weak scale through RG running. Thus, we will have an
expression for $\mu^2(M_Z)$ and $B(M_Z)$ as a function of $\tan \beta$ for
every set of input parameters. Then we scale these expressions back to the
GUT-scale and require that eq. \ref{constr2} is satisfied. This will yield
an equation for $\tan \beta$. In our analysis we use the one loop RGE's
for the MSSM, retaining only the top Yukawa coupling and the gauge
couplings. In this case the RGE's can be solved analytically with the
exception of one function, where numerical integration is necessary. The
approximate analytical expressions are \cite{Ibanez,CW}:

\begin{eqnarray}
&& m_u^2=m_0^2+0.52M_{1/2}^2+\Delta m^2 \nonumber \\
&& m_d^2=m_0^2+0.52M_{1/2}^2 \nonumber \\
&& \mu^2 (M_Z)=2\mu_0^2 \left( 1-\frac{Y_t}{Y_f}\right)^{1/2} \nonumber \\
&& B(M_Z)=B_0-\frac{A_0}{2}\frac{Y_t}{Y_f}+M_{1/2}\left( 1.2\frac{Y_t}{Y_f}
-0.6 \right) ,
\end{eqnarray}
where
\begin{equation}
\Delta m^2=-\frac{3}{2}m_0^2\frac{Y_t}{Y_f}+2.3A_0M_{1/2}\frac{Y_t}{Y_f}
\left( 1-\frac{Y_t}{Y_f}\right)
-\frac{A_0^2}{2}\frac{Y_t}{Y_f}\left( 1-\frac{Y_t}{Y_f}\right) +
M_{1/2}^2\left[ -7\frac{Y_t}{Y_f}+3\left( \frac{Y_t}{Y_f}
\right)^2 \right] ,
\nonumber \end{equation}
\begin{equation}
Y_t=\lambda_t^2/4\pi ,
\nonumber \end{equation}
\begin{equation}
Y_t=\frac{2\pi Y_t(M_G)E(t)}{2\pi +3Y_t(M_G)F(t)},
\nonumber \end{equation}
\begin{equation}
E(t)=(1+\beta_3 t)^{\frac{16}{3b_3}}(1+\beta_2 t)^{\frac{3}{b_2}}
(1+\beta_1 t)^{\frac{13}{9b_1}},
\nonumber \end{equation}
\begin{equation}
\beta_i=\alpha_i(M_G)b_i/4\pi,
\nonumber  \end{equation}
\begin{equation}
t=\log (M_G/\Lambda)^2,
\nonumber \end{equation}
\begin{equation}
F(t)=\int_0^t E(t')dt',
\nonumber \end{equation}
\begin{equation}
Y_f=\frac{2\pi E(t)}{3F(t)},
\end{equation}
for $t=\log (M_G/M_Z)^2$, $M_G=2\;\cdot 10^{16}$
 one has $E\simeq 14, F\simeq 293$ in the MSSM.
Putting these together with the constraint eq. \ref{constr2} and with
eq. \ref{ewbr} for electroweak breaking one gets the following
equation for $\tan \beta$:

\begin{eqnarray}
\label{theeq}
& & m_0^2+\frac{1}{2(1-\frac{Y_t}{Y_f})^{1/2}}\left(
-m_0^2-0.52M_{1/2}^2
-\Delta m^2 \frac{\tan^2 \beta}{\tan^2 \beta -1}-\frac{M_Z^2}{2} \right) =
\nonumber \\ & &
-\frac{1}{\sqrt{2}(1-\frac{Y_t}{Y_f})^{1/4}} \left\{
 - \frac{\tan \beta}{1+\tan^2
\beta}\left( \Delta m^2\frac{\tan^2 \beta +1}{\tan^2 \beta -1}+M_Z^2
\right) \right.
\nonumber \\ & &
\pm \left[ \frac{A_0}{2} \frac{Y_t}{Y_f}-M_{1/2}\left( 1.2\frac{Y_t}{Y_f}
-0.6\right) \right]
\left( -m_0^2-0.52M_{1/2}^2 \right. \nonumber \\ & &
\left. \left. -\Delta m^2\frac{\tan^2 \beta}{\tan^2 \beta -1} -
\frac{M_Z^2}{2}\right)^{1/2}\right\}.
\end{eqnarray}
When solving this equation one has to be careful about the
sign of $\tan \beta$. In the MSSM one can fix $\tan \beta$ to
be positive, because for negative
$\tan \beta$ one can
redefine the phase of one of the Higgs fields to absorb a
minus sign and thus changing the sign of the $B\mu $ term in eq.
2.10. However in our case the constraint of eq. \ref{constr}
is not invariant under this phase redefinition. As a result
eq. \ref{theeq} is not invariant under $\tan \beta \to -\tan
\beta$, thus as opposed to the MSSM one loses generality
by restricting to positive $\tan \beta$. Therefore we
solve eq. \ref{theeq} separately for positive and negative
$\tan \beta$.

The $\pm$ in eq. \ref{theeq}
stands for the two possible signs of the $\mu$ parameter.
Note however, that eq. \ref{theeq}
 does not yet include the corrections to
$m_u^2$ and $m_d^2$ arising from the one loop corrections to the effective
potential which are known to be significant and
which should be incorporated.
The expressions for these corrections are:
\begin{eqnarray}
&& \Delta m_u^2=\frac{\partial \Delta V}{\partial v_u^2}, \nonumber \\
&& \Delta m_d^2=\frac{\partial \Delta V}{\partial v_d^2},
\end{eqnarray}
and we retain only the top-stop loops for $\Delta V$:
\begin{equation}
\Delta V=\frac{3}{16\pi^2}\left[ \frac{1}{2} m_{\tilde{t}_1}^4
\left( \log \left( \frac{m_{\tilde{t}_1}^2}{\Lambda^2}\right) -\frac{3}{2}
\right) +
\frac{1}{2} m_{\tilde{t}_2}^4
\left( \log \left( \frac{m_{\tilde{t}_2}^2}{\Lambda^2}\right) -\frac{3}{2}
\right) -
 m_t^4
\left( \log \left( \frac{m_t^2}{\Lambda^2}\right) -\frac{3}{2}\right) \right] ,
\end{equation}
where $m_{\tilde{t}_{1,2}}$ are the stop masses and $m_t$ is the top mass.
Eq. \ref{theeq} is then modified to be
\begin{eqnarray}
\label{theeq2}
& &
m_0^2+\frac{1}{2(1-\frac{Y_t}{Y_f})^{1/2}}\left( -m_0^2-0.52M_{1/2}^2
-\Delta m^2 \frac{\tan^2 \beta}{\tan^2 \beta -1}-\frac{M_Z^2}{2}+
\frac{\Delta m_d^2}{\tan^2 \beta-1} \right. \nonumber \\ & &
\left. -\frac{\Delta m_u^2\tan^2 \beta}{
\tan^2 \beta -1} \right) =
-\frac{1}{\sqrt{2}(1-\frac{Y_t}{Y_f})^{1/4}}\left\{
-\frac{\tan \beta}{1+\tan^2
\beta}\left[ (\Delta m^2+\Delta m_u^2-\Delta m_d^2 )
\frac{\tan^2 \beta +1}{\tan^2 \beta -1} \right. \right.
\nonumber \\ & & \left. +M_Z^2\right]
\pm \left[ \frac{A_0}{2} \frac{Y_t}{Y_f}-M_{1/2}\left(
1.2\frac{Y_t}{Y_f}-0.6\right) \right]
\left( -m_0^2-0.52M_{1/2}^2-\Delta m^2\frac{\tan^2 \beta}{\tan^2 \beta -1} -
\frac{M_Z^2}{2}+\nonumber \right. \\ & & \left. \left.
\frac{\Delta m_d^2}{\tan^2 \beta-1}-\frac{\Delta m_u^2\tan^2 \beta}{
\tan^2 \beta -1}\right)^{1/2}\right\} .
\end{eqnarray}

\mysection{Results}

We have shown in the previous section that the fact that the $\mu$-term is
generated by the Higgs as PGB mechanism implies eq. \ref{theeq} for
$\tan \beta$.

We note that eq. \ref{theeq} is invariant
under the transformation $\mu \rightarrow -\mu$, $M_{1/2}
\rightarrow -M_{1/2}$ and $A_0 \rightarrow -A_0$. Therefore one
can fix $\mu < 0$, and then the solutions corresponding to
positive $\mu$ are obtained by taking $M_{1/2}\rightarrow
-M_{1/2}$ and $A_0\rightarrow -A_0$. Consequently
we will show four plots;
the first two corresponding to $\tan \beta >0$ and both
signs for $M_{1/2}$ (Figs. 1.c and 1.d). Subsequently we will
give the two plots corresponding to $\tan \beta <0$, with
$M_{1/2}>0$ in Fig. 2 and $M_{1/2}<0$ in Fig. 3. In all these
plots we will have $\mu <0$ fixed. All $\mu >0$ solutions can be obtained
as described above.

The values of $m_0$ and $|M_{1/2}|$ are bounded from below in order to
ensure that the sparticle masses obey the experimental limits \cite{Explim}.
 Thus one has to combine  the experimental lower bounds on the sparticle
 masses and the
requirement that there is a solution to eq. \ref{theeq2}
to get the possible parameter range of the
MSSM.

We solved equation \ref{theeq2} numerically for the small
$\tan \beta$ regime (we take $1< \tan \beta < 15$;
large $\tan \beta$ would require a fine tuning of order $1/\tan \beta$ in
the Higgs sector \cite{NR} contrary to the spirit of not
tuning parameters). In the following we summarize the main
features of the solutions.

\noindent If one fixes $\tan \beta >0$ then the
only viable solutions fulfilling the experimental constraints
correspond to

-large values of $|A_0|$

-relatively small values of $m_0$ and $|M_{1/2}|$.

\noindent This is illustrated in Fig. 1 where we display the allowed
parameter space for a fixed value of $\lambda_t$ and different
fixed values of $m_0$. One gets the same type of plots for
other fixed values of $\lambda_t$. Figs. 1. a and b are presented
to ease the reading of the subsequent plots. They both display
the $m_0=60$ case with detailed explanation of the allowed
parameter space. Figs. 1. c and d are the same but for
four different values of $m_0$.

However there is an important potential
problem with these solutions:
the existence
of charge and/or color breaking (CCB) minima of the potential
of the sleptons and squarks. We use the ``traditional'' condition
for the absence of such minima
\cite{CCB1}  but  we evaluate this condition
 not at the GUT scale
 but at the weak scale. This yields
the following conditions for the weak scale parameters:

\begin{eqnarray}
\label{CCB}
&& A_e^2<3(m_{\tilde{L}}^2+m_{\tilde{e}}^2+m_d^2) \nonumber \\
&& A_d^2<3(m_{\tilde{Q}}^2+m_{\tilde{d}}^2+m_d^2) \nonumber \\
&& A_u^2<3(m_{\tilde{Q}}^2+m_{\tilde{u}}^2+m_u^2),
\end{eqnarray}
where the $A_e, A_d$ and $A_u$ refer to the soft breaking trilinear terms
of a given interaction (that is $A_0$ scaled down to the weak scale)
and the m's refer to the soft breaking scalar mass terms also
evaluated at the weak scale. We have calculated these parameters
in the same one loop approximation (that is including only
loops with gauge or top Yukawa couplings), and plotted the
allowed region (that is the region where the inequalities
\ref{CCB} are satisfied). It is well known that these conditions
are neither sufficient nor necessary to avoid the presence of
CCB vacua \cite{CCB1}. The full analysis for the absence of these
particular CCB vacua has
been recently
done in \cite{CCB2}.
 We have checked that these results are sufficently close
to those of the full analysis given in ref. \cite{CCB2}, with
the results of
ref. \cite{CCB2} being always even more restrictive than those
based on our analysis.

Furthermore even if a CCB vacuum exists that
is a global minimum of the scalar potential one has to
calculate the tunneling rate from the false MSSM vacuum
to the real CCB vacuum for each such solution and only
if that is large can one exclude a given point in the parameter
space. Thus it is clear that eq. \ref{CCB} is not the full
story. However it can be used as an approximate indicator for
the presence of CCB vacua. If one is very far
outside the allowed region allowed by \ref{CCB} then
that
point on the parameter space can be safely excluded.
If one is deep inside the
allowed region CCB vacua are probably not dangerous.

In Fig. 1 we also display the CCB bounds obtained from
eqs. \ref{CCB}, along with the allowed
parameter space. As one can see
from Fig. 1, all these solutions to
equation \ref{theeq2} lay outside the bounds of \ref{CCB} and
thus CCB  poses a threat to the entire allowed parameter space.
Therefore if we take these CCB bounds seriously we have to
discard these solutions. This conclusion is not altered if
we take different values of $\lambda_t$.

\begin{figure}
\begin{center}
\PSbox{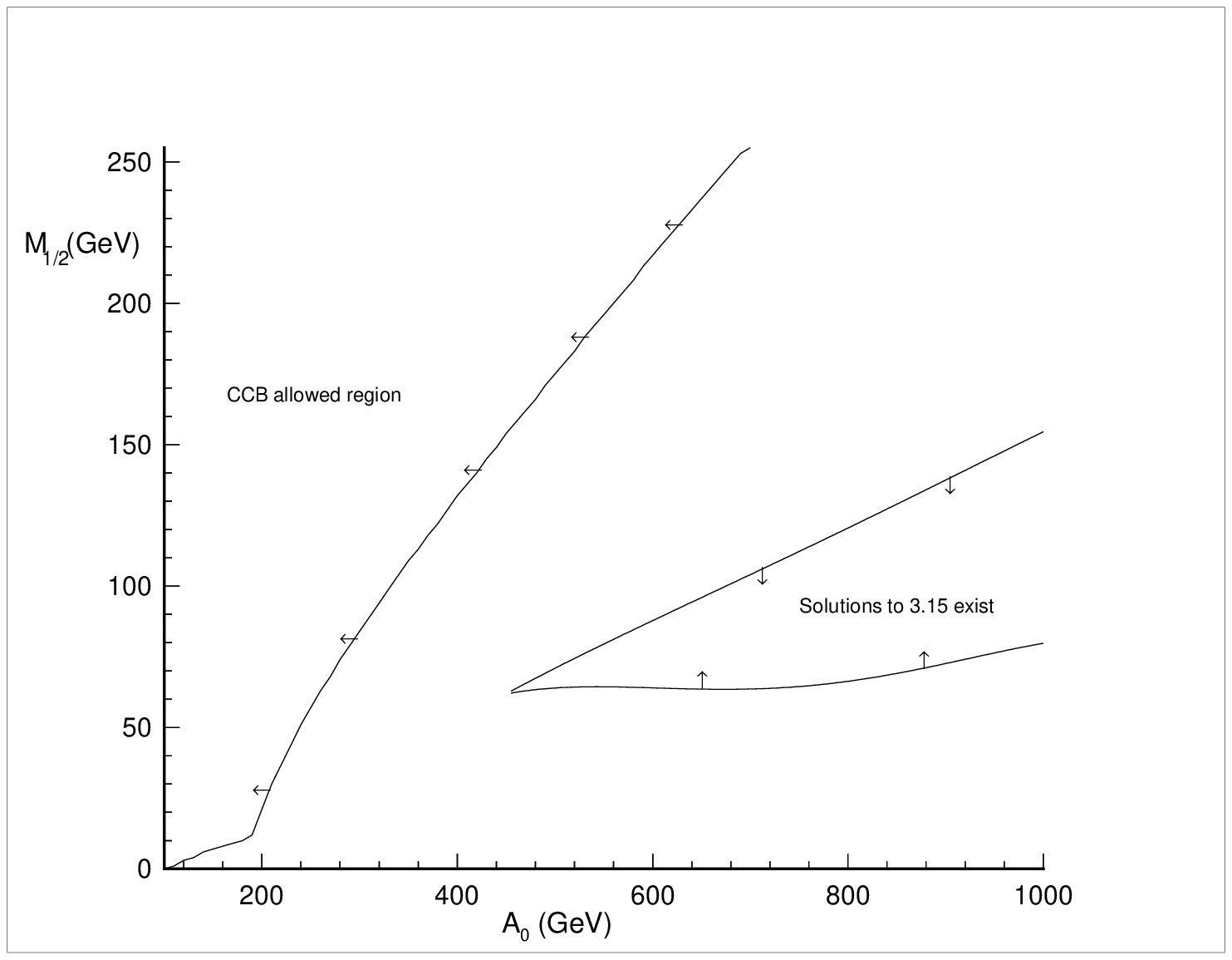 hoffset=-85 voffset=-40
hscale=40 vscale=40}{5cm}{5cm}
\PSbox{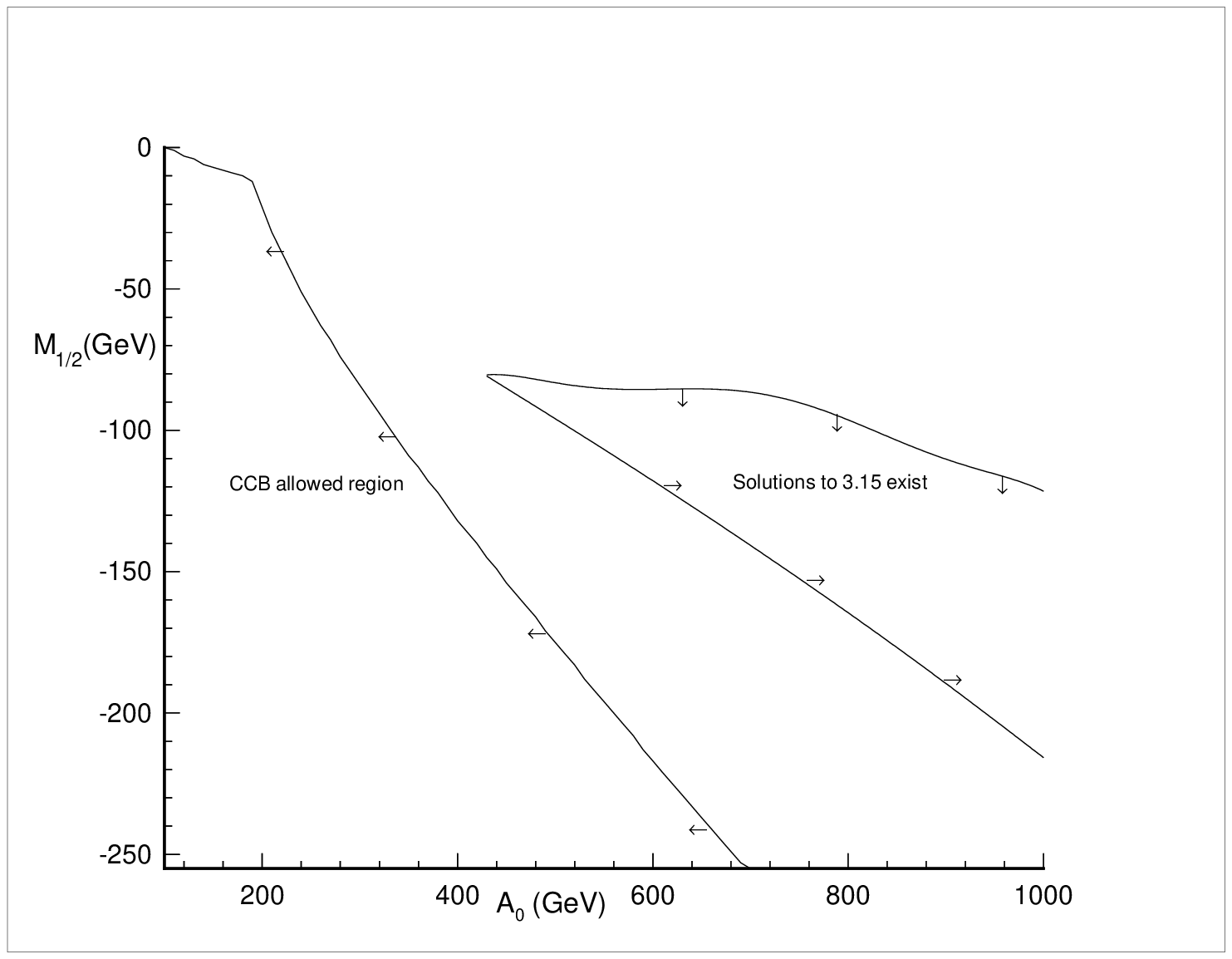 hoffset=-10 voffset=-40
hscale=40 vscale=40}{5cm}{5cm}
\end{center}
\begin{center}
{\figfont \hspace*{3.5cm} Fig. 1.a \hfill Fig. 1.b
\hspace*{3.5cm}}
\end{center}
\begin{center}
\PSbox{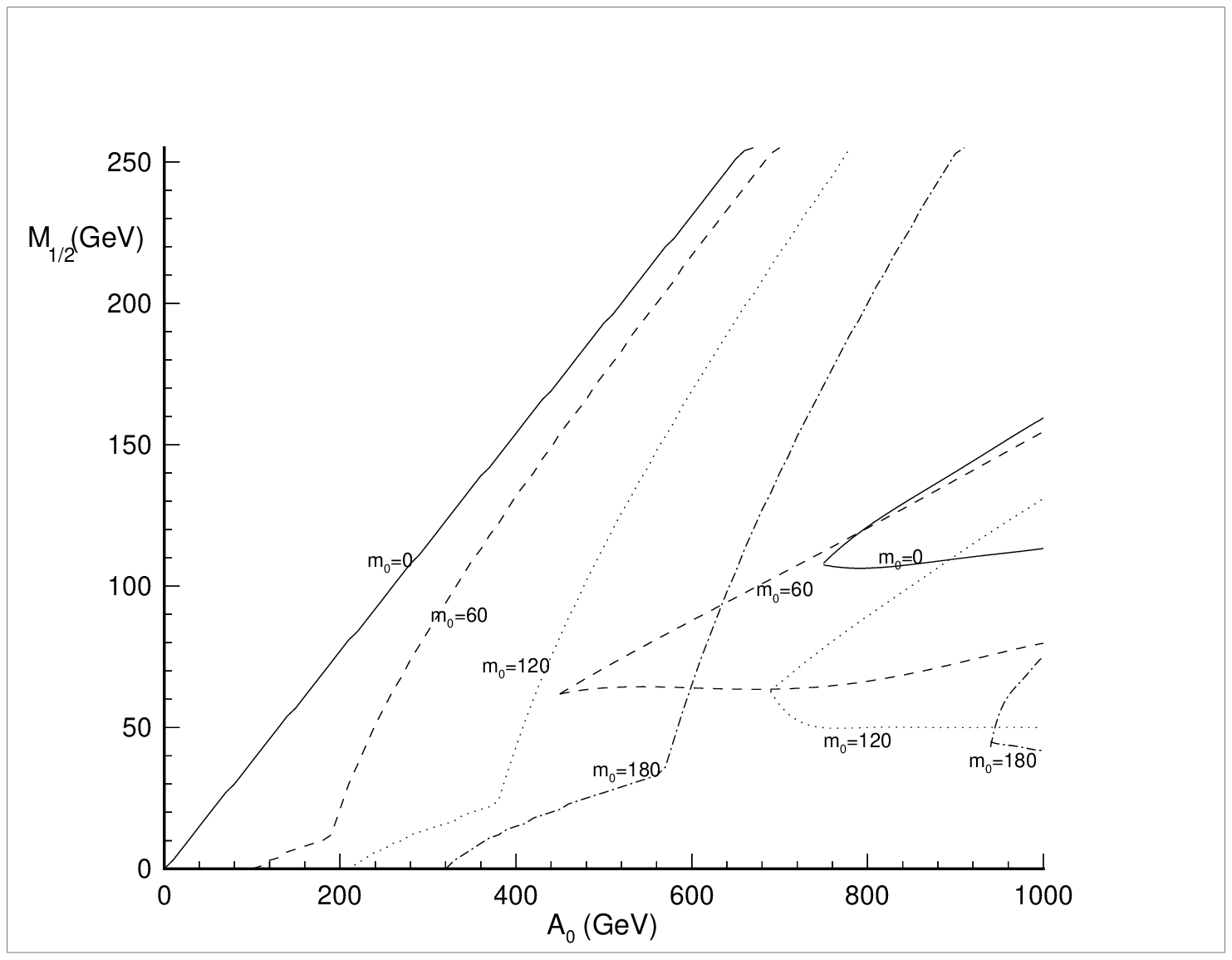 hoffset=-85 voffset=-40
hscale=40 vscale=40}{5cm}{5cm}
\PSbox{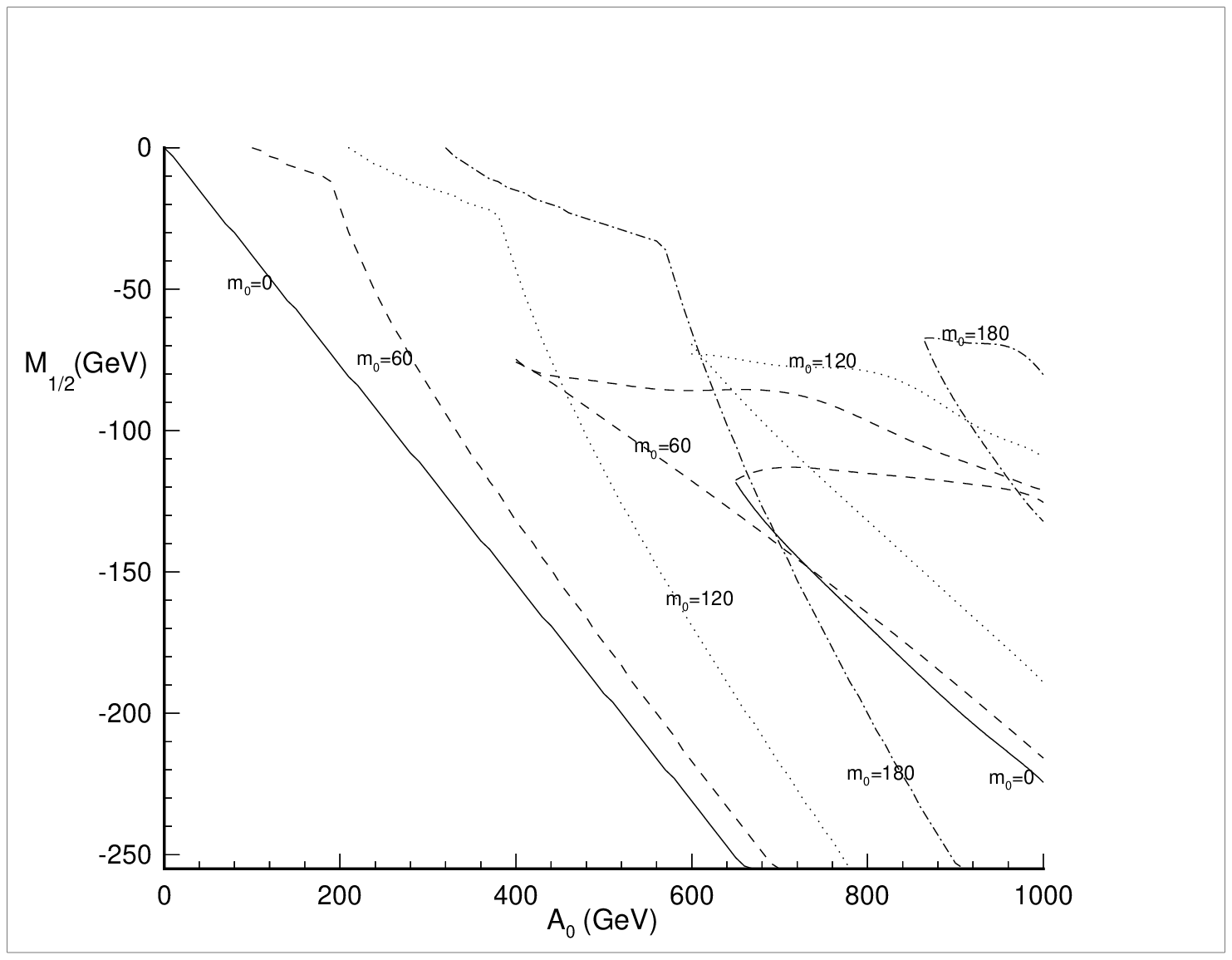 hoffset=-10 voffset=-40 hscale=40 vscale=40}{5cm}{5cm}
\end{center}
{\figfont \hspace*{3.5cm} Fig. 1.c \hfill Fig. 1.d
\hspace*{3.5cm}}


{\figfont Figure 1: The allowed MSSM parameter space for a
fixed value of
{\small $\lambda_t$} {\small $(\lambda_t=1.2)$} and
for different fixed values of {\small $m_0$} with
{\small $\tan \beta >0$}.

a: {\small $m_0=60$
GeV}, {\small $M_{1/2} > 0$}. The curve on the right of the
plot gives the region where equation 3.15 can be
satisfied with the above mentioned parameters, while the
curve on the left gives the region where CCB vacua are absent.

b: the same as in 1.a for {\small $M_{1/2}<0$}.
In both cases {\small $\mu <0$}.
As explained in the text the {\small $\mu >0$} solutions can
be obtained by taking {\small $M_{1/2}\rightarrow -M_{1/2}$} and
{\small $A_0 \rightarrow -A_0$} simultaneously.

c: As in 1.a for varying {\small $m_0$}.
In both Figs. 1.c and 1.d the
solid line corresponds to {\small $m_0=0$ GeV}, the dashed
to {\small $m_0=60$ GeV}, the dotted to {\small $m_0=120$
GeV} and the  dash-dotted to {\small $m_0=180$ GeV}.

d: As in 1.b for varying {\small $m_0$}.}

\end{figure}

In this case we can
conclude that one needs to take $\tan \beta <0$.
If we again fix $\mu<0$ as before. We
will get two type of solutions for this case. The solutions to
the
$M_{1/2}>0$ case resemble very much the solutions of the previous
case; that is  one needs to have large $|A_0|$ and small
$m_0$ and $M_{1/2}$. However now larger values of $M_{1/2}$
and $m_0$ are possible, but the overlap with the CCB allowed
region is still very small as illustrated in Fig. 2.
(There is still no overlap for small values of $m_0$'s and tiny
overlap for large values.)

\begin{figure}
\begin{center}
\PSbox{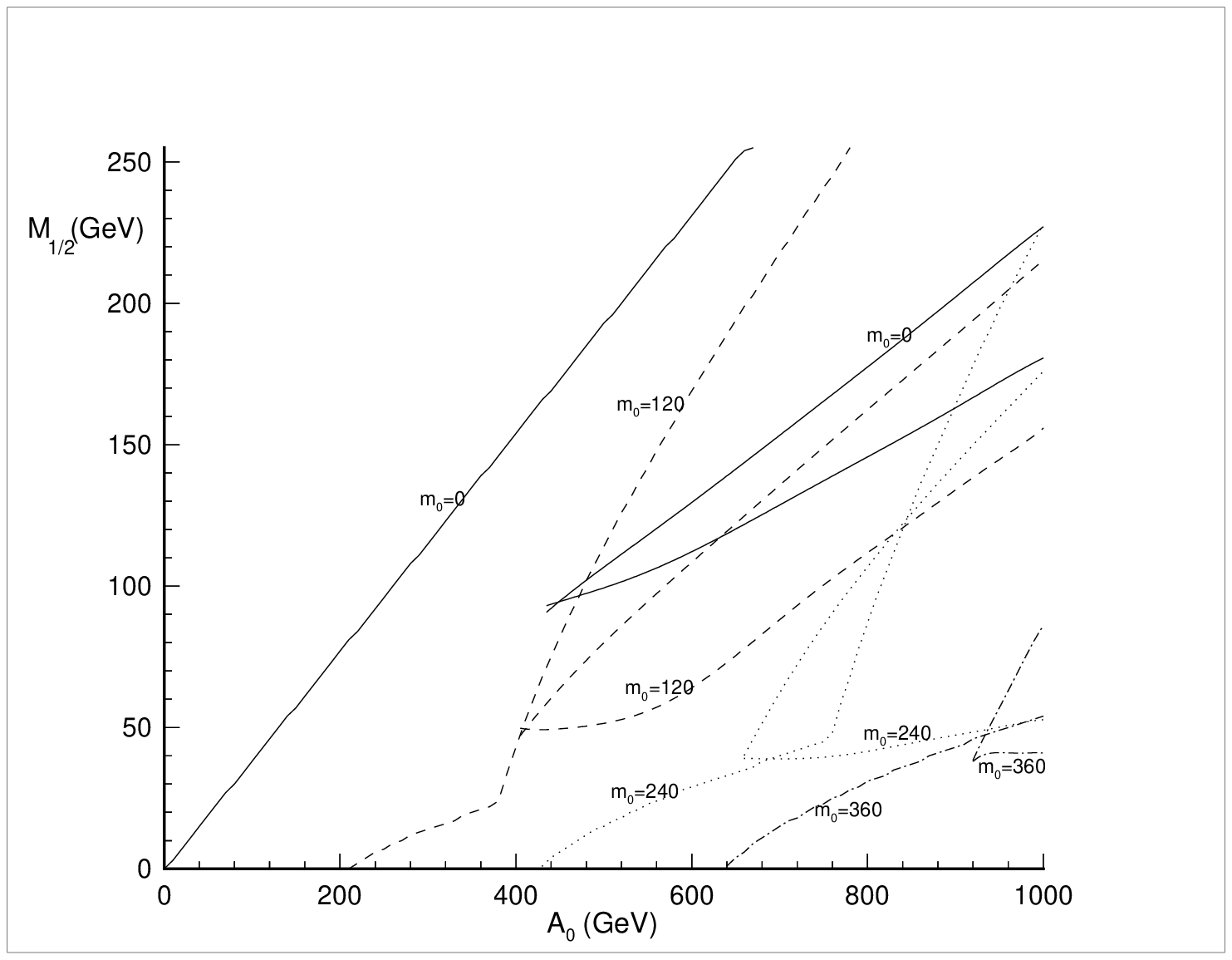 hoffset=-85 voffset=-40
hscale=40 vscale=40}{5cm}{5cm}
\end{center}

{\figfont Figure 2:
The same as Fig. 1 but for {\small $\tan \beta <0$}
and positive $M_{1/2}$.
The solid line corresponds to
{\small $m_0=0\; {\rm GeV}$},
 the dashed to {\small $m_0=120\; {\rm  GeV}$},
 dotted to {\small $m_0=240\; {\rm  GeV}$} and
the dash-dotted to {\small $m_0=360\; {\rm GeV}$}.
One can see that the overlap of the solutions with the
CCB allowed regions is very tiny.}
\end{figure}

The final possibility (negative $\tan \beta$, negative $\mu$
and negative $M_{1/2}$) is not so restrictive.
In this case one does not get an upper bound
on $|M_{1/2}|$ and $m_0$; instead one gets a lower bound
on $|M_{1/2}|$ (which is
however more constraining then the usual experimental bounds
in the MSSM). One gets a large overlap with the
CCB allowed region. This region of overlap is however much more
restricted than the region of parameters allowed in the MSSM.
This
is illustrated in Fig. 3 for different values of $m_0$ and
fixed $\lambda_t$. Figs. 3.a and 3.c are again presented to
ease the reading of the other two plots, with detailed
explanation  of the allowed region. Note in Figs. 3.b
and 3.d that for $m_0=500, 1000$ GeV one does not get
any restriction from the CCB bounds if $|A_0|<1000$ GeV. Also
note that for large values of $m_0$ the MSSM bound on $M_{1/2}$
is independent of $m_0$. Different values of $\lambda_t$ yield
similar plots, with the curves somewhat shifted to the right
(towards larger values of $A_0$). This is illustrated in Figs.
3.b and 3.d.

In summary we find
that most of the possible solutions to eq. \ref{theeq2}
obeying the CCB bounds correspond to the $\tan \beta ,
M_{1/2},\mu <0$ case
with the allowed parameter space displayed in
Fig. 3.

\begin{figure}
\begin{center}
\PSbox{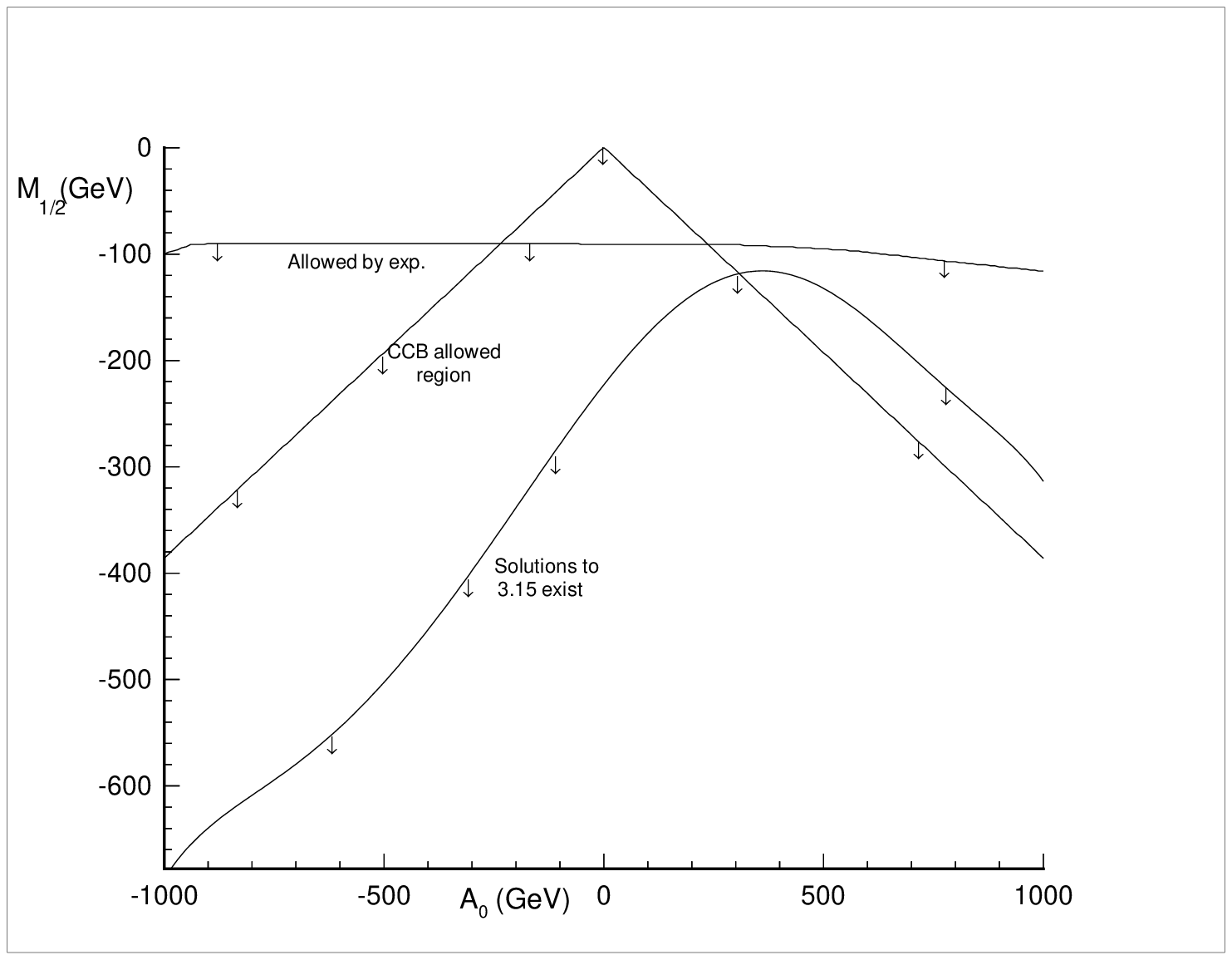 hoffset=-85 voffset=-40
hscale=40 vscale=40}{5cm}{5cm}
\PSbox{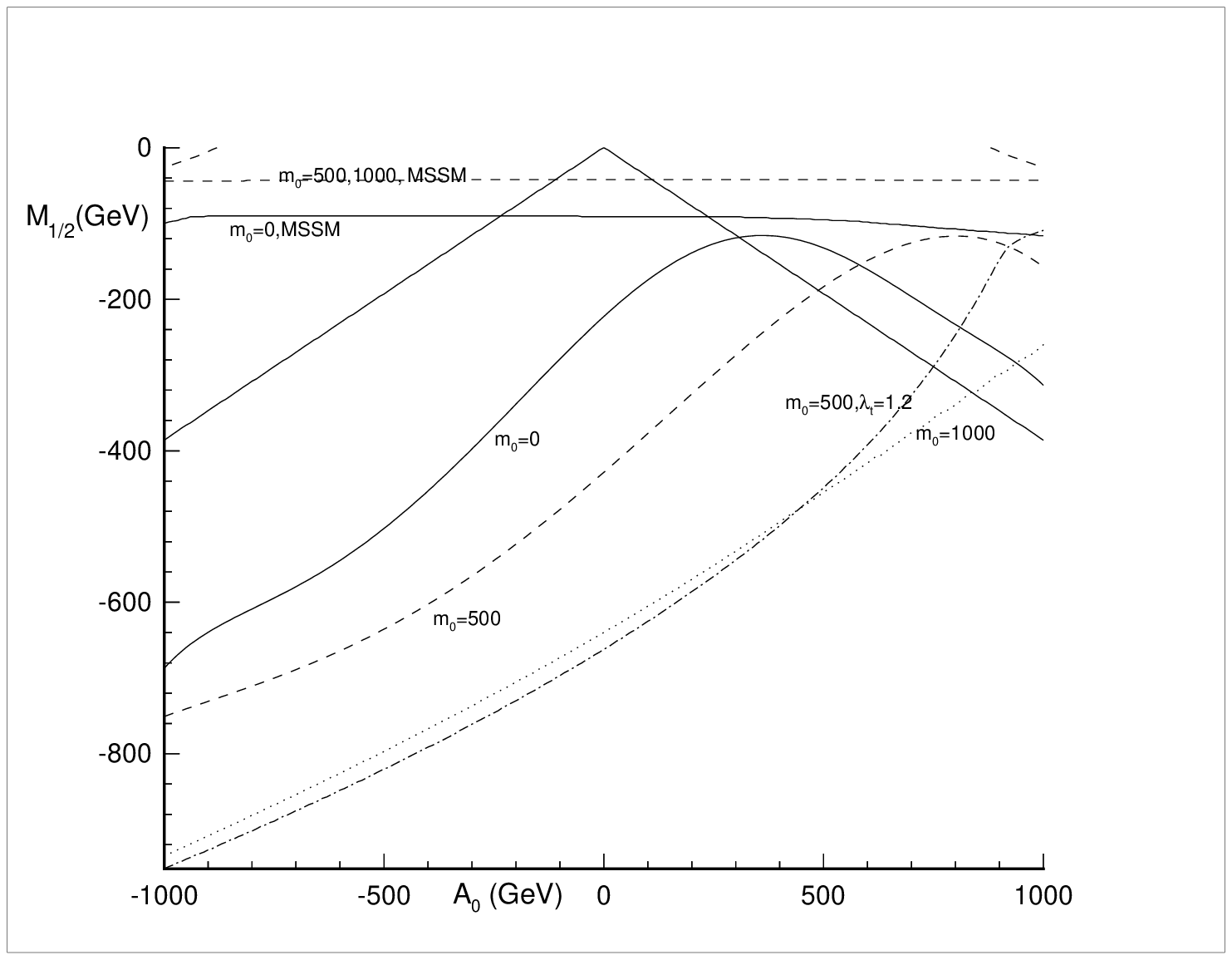 hoffset=-10 voffset=-40
hscale=40 vscale=40}{5cm}{5cm}
\end{center}
{\figfont \hspace*{3.5cm} Fig. 3.a \hfill Fig. 3.b
\hspace*{3.5cm}}
\begin{center}
\PSbox{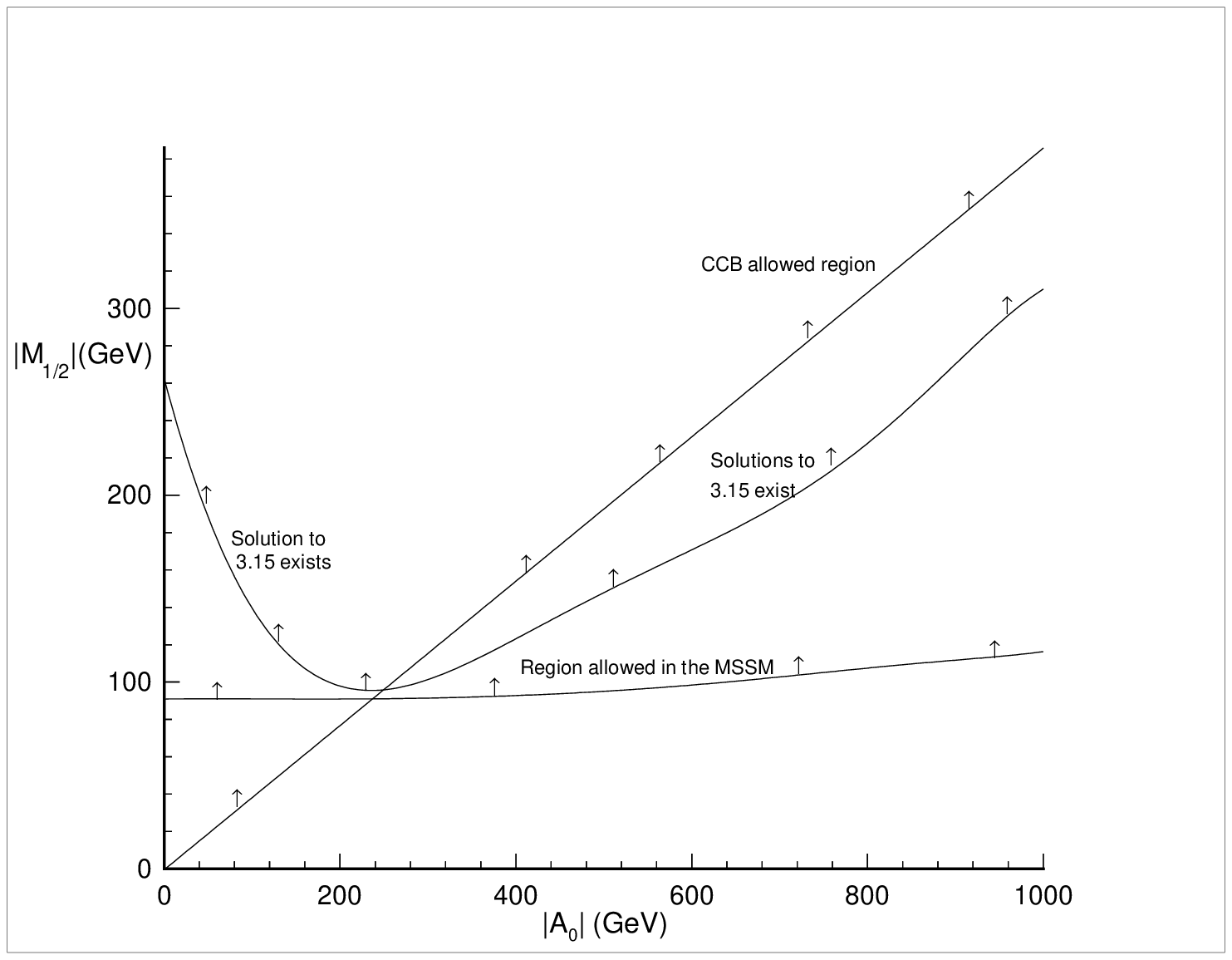 hoffset=-85 voffset=-40
hscale=40 vscale=40}{5cm}{5cm}
\PSbox{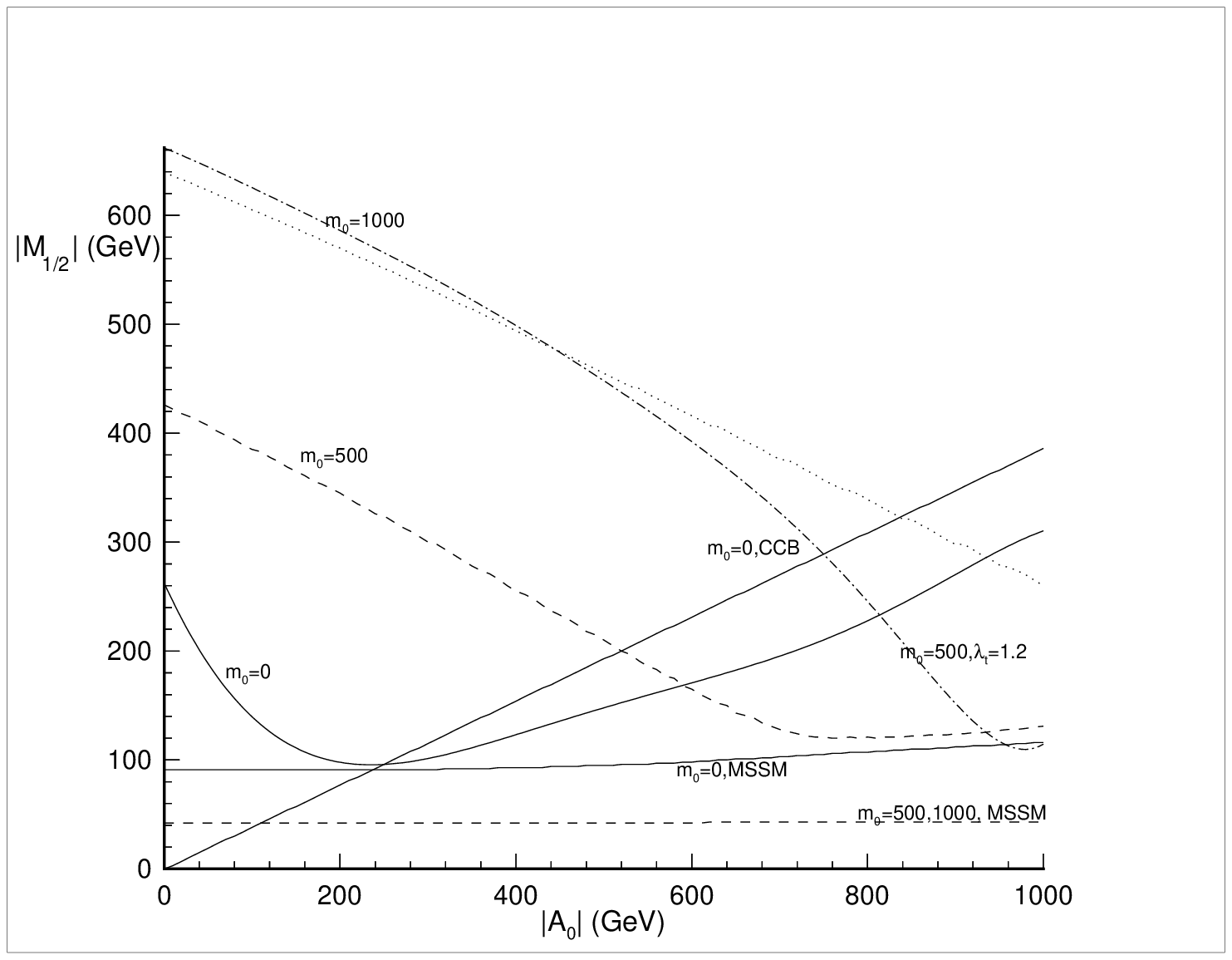 hoffset=-10 voffset=-40
hscale=40 vscale=40}{5cm}{5cm}
\end{center}
{\figfont \hspace*{3.5cm} Fig. 3.c \hfill Fig. 3.d
\hspace*{3.5cm}}

{\figfont Figure 3: The allowed region of parameters
for {\small $\tan \beta <0$}
and negative {\small $M_{1/2}$}. {\small $\lambda_t=
0.8$} for all four plots (except in Figs. 3.b and
3.d where explicitly stated)
 a: {\small $m_0=0$ GeV}. The upper straight line
corresponds to the bound on {\small $M_{1/2}$}
in the MSSM, if the values of {\small
$\tan \beta$} are varied between {\small 1} and {\small 15}.
The {\small $\Lambda$}-shaped curve on the top of the plot gives
the region allowed by the absence of CCB vacua, while the
lower line corresponds to the bound on {\small $M_{1/2}$}
obtained from equation 3.15.
b: The same as in Fig. 3.a but
for different values of {\small $m_0$}.
The solid lines corresponds to {\small $m_0=0$ GeV}, the
dashed lines to {\small $m_0=500$ GeV} and the
dotted line to {\small $m_0=1000$ GeV}. Note that one does not
get any restriction from the CCB bounds for {\small $m_0=500,
1000$ GeV} when {\small $|A_0|<1000$ GeV}.
Also note that for large values of {\small $m_0$} the
MSSM bound on {\small $M_{1/2}$} is independent of
{\small $m_0$}. The dash-dotted line corresponds to
{\small $m_0=500$ GeV}, but with {\small $\lambda_t=1.2$}.
The increase in {\small $\lambda_t$} results in the
shift of the
curves towards larges values of {\small $A_0$}.
c: The bounds on the absolute value of {\small $M_{1/2}$}
as a function of the absolute value of {\small $A_0$}
obtained from Figs.
3.a-b. 3.c gives the {\small $m_0=0$} case,
where the lower straight curve is the MSSM bound, the upper
straight curve is the bound from CCB, while the third curve
in the middle is the bound obtained by requiring that
3.15 has a viable solution.
d: The same as in c, where {\small $m_0=0$ GeV}
 corresponds to the
solid lines, {\small $m_0=500$ GeV} to the dashed lines and
{\small $m_0=1000$ GeV} to the dotted line. The dash-dotted
line corresponds to {\small $m_0=500$ GeV} but
{\small $\lambda_t=1.2$}.}

\end{figure}

Since
$\tan \beta$ is not a free parameter of the theory it is not
surprising that the range of values $\tan \beta$ can take on
is much smaller than in the MSSM.
There  any low value of $\tan \beta $
not too close to 1 can be acceptable for fixed $A_0$ and $m_0$
if one varies $M_{1/2}$. In our case however $\tan \beta$ is
the solution to eq. \ref{theeq2} and thus will in general not
take all values. This is illustrated in Fig. 4, where we
display the allowed range of $\tan \beta$ for fixed values
of $\lambda_t$ and $m_0$, while $|M_{1/2}|$ is allowed to vary
in the range of $0,800$ GeV. One can see that one finds a
smaller region of acceptable vacua
than the MSSM together with the experimental
constraints would allow.

To conclude this section we summarize the consequences
of our analysis.
We have seen that the boundary condition \ref{constr}
together with the requirement of an acceptable SM minimum
will determine $\tan \beta$ from other input parameters. We
have seen that this equation for $\tan \beta$ does not always have
solutions which excludes some regions of the MSSM parameter
space (which is now reduced to $m_0,\; M_{1/2}, \; A_0, \;
\lambda_t$). These constraints displayed in Figs. 3 and 4
are the main results of our analysis. If the $\mu$-term is
generated by the Higgs as PGB mechanism then the MSSM
parameters must be inside the boundaries given in Figs. 3 and 4
or inside the tiny overlapping regions of Fig. 2.

Thus if one ultimately measures these parameters
in colliders one can check whether they are indeed in the allowed
region or not. If the MSSM parameters are all measured one
can also check whether the experimental value of
$\tan \beta$  does satisfy eq. \ref{theeq} or not, thereby
testing the assumptions on the GUT-scale Higgs sector.

\begin{figure}
\begin{center}
\PSbox{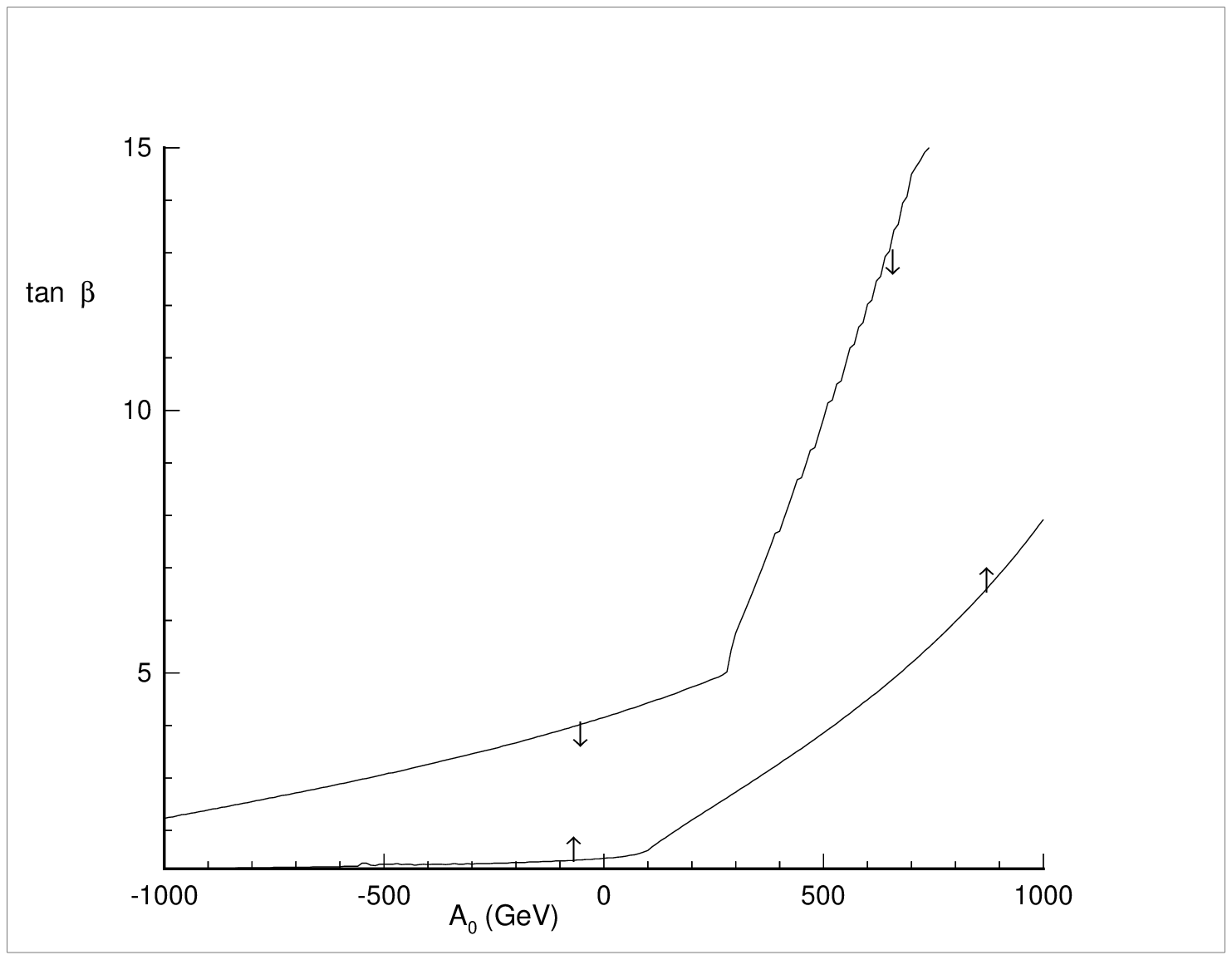 hoffset=-37 voffset=-40
hscale=30 vscale=30}{5cm}{5cm}
\PSbox{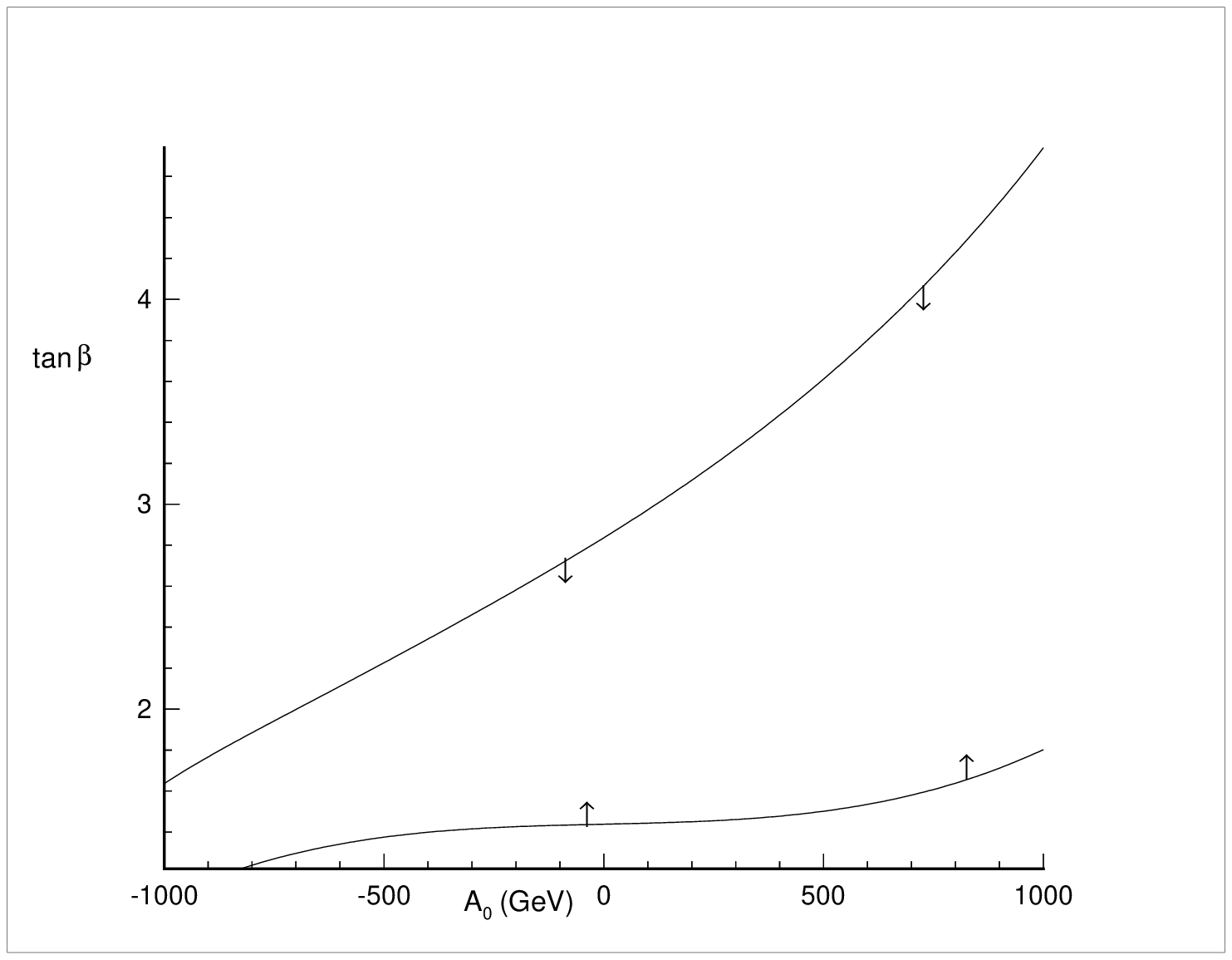 hoffset=-12 voffset=-40
hscale=30 vscale=30}{5cm}{5cm}
\PSbox{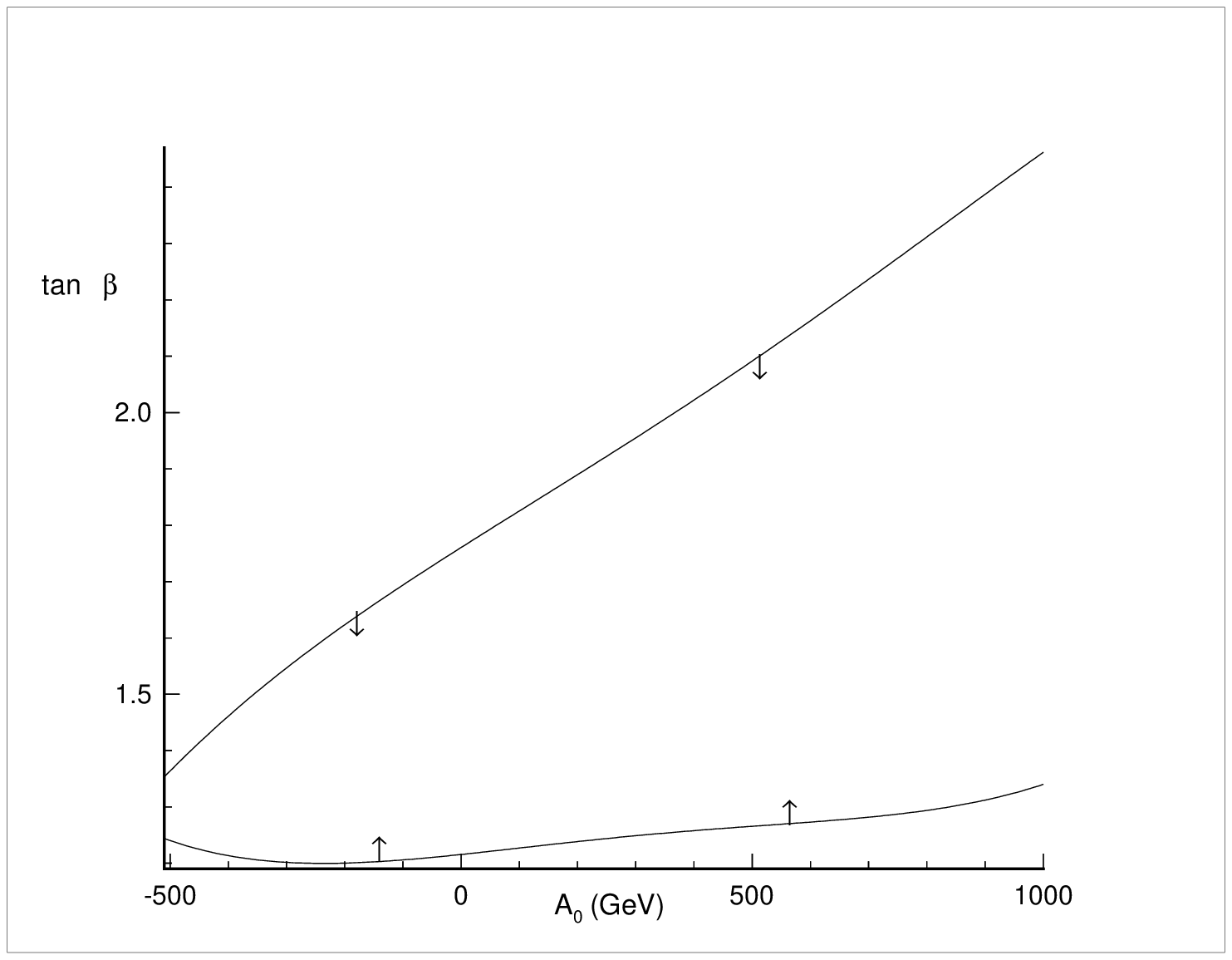 hoffset=13 voffset=-40
hscale=30 vscale=30}{5cm}{5cm}
\end{center}
\hspace*{1cm}
\vspace*{-1cm}
{\figfont \hspace*{1cm} Fig. 4.a \hspace{4cm} Fig. 4.b
\hfill Fig. 4.c  \hspace*{1cm}}

\vspace*{1cm}

{\figfont Figure 4: a: The allowed region for {\small $\tan
\beta$} if we vary {\small $|M_{1/2}|$} between {\small $0$}
and {\small $800$ GeV}. {\small $m_0=0$ GeV} and {\small
$\lambda_t=0.8$} is fixed. b and c are the same as a
with {\small $m_0=500, \; 1000$ GeV}.}
\end{figure}

\mysection{Implications for realistic models}

We finally comment on the validity of our analysis
for models implementing the Higgs as PGB scheme.
Although
 the Higgs as pseudo-Goldstone boson idea
is perhaps   the most natural solution to the
doublet-triplet splitting problem in the
context of SUSY GUT's, it is   difficult to build
realistic models that implement this idea in a natural way (see ref.
\cite{BCR}) without additional light charged particles which
disrupt unification.

The only known realistic model is based on the SU(6) gauge
group \cite{Zur,Bar2,Bar3,Zur2,BCR}
and has an accidental SU(6)$\otimes $SU(6) global
symmetry of the Higgs sector. This symmetry is achieved by
requiring that two sectors of the Higgs fields are are not mixed
among each other in the superpotential. The models of refs.
\cite{Zur,Bar2,Bar3,Zur2,BCR}
use the SU(6) adjoint $\Sigma$ and a pair of SU(6)
vectors $H,\bar{H}$ for the Higgs sector. Then the superpotential
has the form
\begin{equation}
W(\Sigma ,H,\bar{H})=W_1(\Sigma )+W_2(H,\bar{H})
\end{equation}
up to dimension seven in the superpotential of the Higgs fields.
This stringent requirement is necessary so that
nonrenormalizable
operators breaking the accidental global symmetry do not give
too large a mass to the Higgs doublets. In ref. \cite{BCR}
several suggestions for a superpotential implementing this idea
have been
presented. From the point of view of the $\mu$-term we can
divide them into two categories, according to whether a
symmetry breaking term (that is a term that couples the
$\Sigma$ and $H,\bar{H}$ fields)
containing seven Higgs sector fields is or is not allowed.\footnote{The models
 in ref. \cite{BCR} were
especially designed such that no
symmetry breaking terms containing only six or less Higgs sector fields
are allowed, since these would give the
doublet Higgses a mass of order $M_{GUT}
(\frac{M_{GUT}}{M_{Pl}})^{-3}\sim 10^{7}$ GeV
and thus
spoil the solution to the
doublet-triplet splitting problem.} A dimension
seven operator would give an additional contribution to
the $\mu$-term spoiling eq. \ref{theeq2} without destroying
the solution to the doublet-triplet splitting problem.

In model 2 of ref. \cite{BCR} such a term is allowed by
all symmetries of the Lagrangian and thus may yield a
contribution to the $\mu$-term of order 100-1000 GeV. Since the
coefficient of this operator is a completely free parameter of
the theory the constraint of eq. \ref{theeq2} does not hold
and our analysis may not be applied to this theory. Such
a theory cannot be tested by the constraints described
in this paper.

However, if
 no  dimension seven mixing terms are allowed in the
superpotential then there can be
no significant extra contribution to the $\mu$-term. This is the case
in the simplest model, namely model 1 of ref. \cite{BCR}
and also in model 3 of the same reference.

The superpotential of model 1 of ref. \cite{BCR}
is given by
\begin{equation}
\frac{1}{2} M {\rm Tr} \Sigma^2 +\frac{1}{3} \lambda
{\rm Tr} \Sigma^3 +\frac{\alpha}{M_{Pl}^{2n-3}} (\bar{H}H)^n,
\end{equation}
where n=4,5,6. After the inclusion of the soft breaking
terms one gets $\langle H \rangle \sim 10^{17} \; {\rm GeV}
>M_{GUT}$, and at this scale SU(6) is broken to SU(5). If one
neglects the small admixture of $H,\bar{H}$ fields in the
Higgs doublets then at the SU(5) scale $\langle H\rangle$
we have an SU(5) gauge theory with an ``accidental'' global
SU(6) symmetry of the Higgs sector, since the
theory originates from an SU(6) gauge theory.
 Thus at the $\langle H
\rangle$ scale we get as an effective theory exactly the
model of Section 2 since the SU(5) nonsinglet fields of
$H$ are eaten by the heavy gauge bosons.
This means that the threshold corrections to eq. \ref{constr}
arising from the fact that the constraint is not generated at
the GUT scale but at a somewhat higher scale
can be estimated
to be
of the order
\begin{equation}
\frac{\lambda_t^2}{16\pi^2} \log \left( \frac{\langle H \rangle}{M_{GUT}}
\right) \sim 0.01,
\end{equation}
due to the running between the $\langle H \rangle$ and GUT
scales.
Thus the corrections in this model
to eq. \ref{theeq} are expected to be a
few percent and the results of our analysis should not
be modified significantly. We have checked that
corrections in eq. \ref{constr} as large as 10 percent
caused only a small
shift in the constraint curves. Consequently there
was still
no overlap between
the allowed parameter region and the region allowed by CCB for
the $\tan \beta >0$ case.
Therefore
the constraints obtained in this analysis should be robust.

\mysection{Conclusions}

We have investigated the consequences of the assumption that
the supersymmetric
$\mu$-term is entirely due to the Higgs as PGB mechanism.
In this case the number of independent parameters is reduced
by one; thus one can determine $\tan \beta$ as a function of
the other parameters.

We have derived the additional constraint
and examined its consequences. We found that for some region
of input parameters this equation can not be fulfilled,
constraining the possible parameter space. Namely we have found
that one has either large $|A_0|$ and small $M_{1/2}$ and $m_0$,
which is disfavored by the potentially dangerous presence of
CCB vacua, or we get an ($m_0$ and $A_0$ dependent) lower
bound on $|M_{1/2}|$.  In addition since $\tan \beta$ is
determined by the other parameters,
the allowed range of $\tan \beta$ is also constrained. Further
experimental consequences might be revealed in subsequent
probes of the restricted parameter space.

These bounds are consequences of the assumptions about the
Higgs sector of the underlying
GUT theory. Thus if SUSY is discovered, the
measurement of the low energy parameters should also serve
as a probe of the GUT scale physics.

\section*{Acknowledgements}

We are grateful to
Diego Casta\~no for many
 helpful suggestions and to Marcela Carena, Alex Kusenko
and Carlos Wagner
for useful  discussions.

\end{document}